
\input phyzzx


\catcode`\@=11
\def\binrel@#1{\setbox\z@\hbox{\thinmuskip0mu
 \medmuskip\m@ne mu\thickmuskip\@ne mu$#1\m@th$}%
 \setbox\@ne\hbox{\thinmuskip0mu\medmuskip\m@ne mu\thickmuskip
 \@ne mu${}#1{}\m@th$}%
 \setbox\tw@\hbox{\hskip\wd\@ne\hskip-\wd\z@}}
\def\underset#1\to#2{\binrel@{#2}\ifdim\wd\tw@<\z@
 \mathbin{\mathop{\kern\z@#2}\limits_{#1}}\else\ifdim\wd\tw@>\z@
 \mathrel{\mathop{\kern\z@#2}\limits_{#1}}\else
 {\mathop{\kern\z@#2}\limits_{#1}}{}\fi\fi}
\def\utilde#1{\underset{\widetilde{\ }} \to #1}


\pubnum={214/COSMO$-$28}

\titlepage
\vskip 1cm
\title{ Wave Function of the Universe in Topological and in Einstein
2-form Gravity
\footnote{\dagger}{
Talk given at the Workshop on General Relativity and Gravitation
held at Waseda University, January 18-20, 1993.}}
\vskip 1cm
\author{
\   Akika Nakamichi \footnote{\ast}{e-mail address:
akika@phys.titech.ac.jp}}

\vskip 1cm

\address{
          Department of Physics, Tokyo Institute of Technology
         \nextline Oh-okayama, Meguro-ku, Tokyo 152, Japan
\nextline
}

\vskip 1 cm

\abstract{ We clarify the relation between  2-form Einstein gravity
and its topological version.
 The physical space of the topological version is contained in
that of the Einstein gravity.
 Moreover a new vector field is introduced into 2-form
Einstein gravity to restore the large symmetry of  its topological version.
The wave function of the universe is obtained for each model.
 }

\vfill

\eject

\sequentialequations


\chapter{Introduction }                          %

 Topological gravity is expected to elucidate the global (topological)
aspects of gravity.
 In Ref.[1], a topological version of four dimensional Einstein gravity
is proposed.
 This topological version is obtained by modifying an alternative
formulation of gravity enlightened by Capovilla et al.[2], in which
anti-self-dual 2-forms are used as fundamental variables.
 This formulation, which we call 2-form Einstein gravity,
leads to the canonical formalism discovered by Ashtekar [3].
 Investigating the relation between 2-form Einstein gravity and
its topological version, we found that, a unique quantum state in the
latter turns out to be one of the physical states in the former.
 This physical state is interpreted as the wave funvtion of the
universe.  \par
 Moreover in Ref.[4], a new vector field is introduced into 2-form
Einstein gravity to restore the large symmetry of its topological
version.
 We also obtain the wave function of the universe in the new system.

\chapter{Symmetries}                              %

\subsection{Einstein Gravity}

 The action of (Euclidean) 2-form Einstein gravity is given in terms of
2-form $\Sigma^{k}$ and SU(2) spin connection 1-form $\omega^{k}$ in
the presence of a cosmological constant $\Lambda$,
$$
 S = \int {\Sigma^{k}} \wedge {R_{k}}
  -{\Lambda \over 24} \  {\Sigma^{k}} \wedge {\Sigma_{k}}
  +{\alpha \over 2} \  {\psi_{kl}}{\Sigma^{k}} \wedge {\Sigma^{l}}
  \ ,
                                                            \eqno(1)
$$
where $R_{k} \equiv d\omega_{k} + (\omega \times \omega)_{k} $,
$\, \psi_{kl} $ is a symmetric trace-free Lagrange multiplier field, and
$\alpha$ is a constant parameter.
 The SU(2) indices $i,j,k, \cdots = 1,2,3,$ in the fields
imply that they transform under the ${\it chiral}$ local-Lorentz
representation $(1,0)$ of SU(2)$\times$SU(2) [5].
 In this formulation, the metric $g_{\mu \nu}$ is defined as
\footnote{{}^{\sharp 1}}{We use the notation for the SU(2) indices,
       $F \cdot G \equiv F^i G^i$ and
        $(F \times G)^i \equiv \varepsilon_{ijk} F^j G^k$, where
          $\varepsilon_{ijk}$ is the structure constant of SU(2). }
$$
g^{1\over 2} g_{\mu \nu}
           = -{1 \over 12} \, {\epsilon^{\alpha \beta \gamma \delta}}
                          \, {\Sigma_{\mu \alpha}} \cdot
                             ({\Sigma_{\beta \gamma}} \times
                               {\Sigma_{\delta \nu}}) \ ,
            \qquad g \equiv det(g_{\mu \nu})\ .
                                                            \eqno(2)
$$
 Using this definition, we find that the action (1) is equivalent to
the usual Einstein-Hilbert action [2,6]. \par

 Since the action (1) describes general relativity, it is invariant
under the local-Lorentz transformation and diffeomorphism,
$$
  \delta \omega^{k} = D\theta^{k}_0 + {\cal L}_\xi \omega^{k}\ ,
\qquad
  \delta \Sigma^{k} = [\Sigma, \theta_0]^{k} + {\cal L}_\xi \Sigma^{k}    \ ,
                                                            \eqno(3)
$$
where ${\cal L}_\xi$ is the Lie derivative with respect to a vector
field $\xi^\mu$, and the local-Lorentz transformation corresponds to
the SU(2) gauge transformation with a parameter $\theta^{k}_0$. \par

\subsection{Topological Gravity}

 {}From the equations of motion,
 the multiplier field  $\psi_{kl}$ is determined to be
proportional to the anti-self-dual part of the Weyl tensor,
 which  governs the modes of the gravitational wave.
 Therefore the topological version of the theory is obtained by
simply dropping
the last term in the action (1), that is, by setting $\alpha$ = 0 [1]:
$$
  S_{\alpha =0}= \int \Sigma^k \wedge R_k
                   -{\Lambda \over 24} \ \Sigma^k \wedge \Sigma_k \ .
                                                            \eqno(4)
$$
  In this case, a new symmetry generated by a parameter 1-form
$\theta_1^k$ emerges in addition to diffeomorphism and the local-Lorentz
(with $\theta_0^k$) symmetries,
$$
  \delta \omega^k =D\theta_0^k +{\Lambda \over 12}\theta_1^k  \ ,
\qquad
  \delta \Sigma^k =2(\Sigma \times \theta_0 )^k + D\theta_1^k \ .
                                                              \eqno(5)
$$
 Here diffeomorphism with a vector field $\xi^\mu$ is implicitly
included in the above local-Lorentz and
$\theta_1^k$- transformations
as we can see by setting
$\theta_0^k
= \xi^\nu \omega_\nu^k$ and $\theta_{1 \mu}^k
= 2\xi^\nu \Sigma_{\nu \mu}^k$.
 The theory turns out
to be on-shell reducible in the sense that the transformation laws (5)
are invariant, modulo the equations of motion, under
$$
    \delta \theta_0^k = - {\Lambda \over 12} \epsilon_0^k \ ,
  \qquad
    \delta \theta_1^k = D \epsilon_0^k \ .               \eqno(6)
$$
 This means that not all of the parameters in (5) are independent. \par

\subsection{New System}

  {}From the view point of the topological gravity, one can see
that the large $\theta_1^k$-symmetry is partially broken in Einstein
gravity leaving only diffeomorphism and local-Lorentz symmetries intact
and, as a result, the modes of the gravitational wave are induced.
 The obstruction for the $\theta_1^k$-symmetry is the last term in the
action (1).
 We can restore the symmetry by introducing a vector field (1-form)
$\eta^k$ in the last term as follows,
$$
  {\alpha \over 2} \int \, {\psi_{kl}}{\Sigma^{k}} \wedge {\Sigma^{l}}
\quad
\Rightarrow
\quad
  {\alpha \over 2} \int \, {\psi_{kl}}{\hat \Sigma}^{k} \wedge
                           {\hat \Sigma}^{l} \ ,
\quad
  {\hat \Sigma}^{k} \equiv {\Sigma^{k}}- D {\eta}^{k}+
                {\Lambda \over 12} (\eta \times \eta)^{k} \ .
                                                             \eqno(7)
$$
 This makes our new system invariant under the
$\theta_1^k$-transformation in (5) with $\delta \eta^k = \theta_1^k$,
together with diffeomorphism and the local-Lorentz transformation:
$$
\eqalign{
 & \delta \omega^{k} = D\theta^{k}_0 + {\cal L}_\xi \omega^{k}
                        + {\Lambda \over 12}\theta_1^k  \ ,
\qquad
  \delta \Sigma^{k} = [\Sigma, \theta_0]^{k} + {\cal L}_\xi \Sigma^{k}
                  + D\theta_1^k \ .
\cr
  & \delta \eta^{k} = [\eta, \theta_0]^{k} + {\cal L}_\xi \eta^{k}
                    + \theta_1^k \ .}
                                                            \eqno(8)
$$
 They are all independent and hence there is no reducibility in the
system.
 It is easily shown that the new system is equivalent to Einstein
gravity, by choosing the gauge condition, $\eta^k=0$, for the
$\theta_1^k$-symmetry.
 Indeed physical degrees of freedom of the new system is 2: the modes
of the gravitational wave.

\chapter{Physical states}                         %

\subsection{Topological Gravity}

 In the topological case ($\alpha$ = 0), the action (4) becomes in
canonical form:
$$
   S = \int dt \int d^3x [ \dot \omega_a \cdot B^a
         - \omega _0 \cdot \varphi
            - \Sigma_{a0} \cdot \phi^a ] \ .
                                                            \eqno(9)
$$
 The canonical variables are $\omega_a^k$ and their conjugate momenta
$B^a_k \equiv \varepsilon ^{abc} \Sigma_{bc}^k$,
 which are the spatial components of the spin connection $\omega^k$
and the 2-form $\Sigma^k$.
 Varying (9) with respect to their time components $\omega_0^k$ and
$\Sigma_{a0}^k$, we get two sets of constraints,
$$
   \varphi _k \equiv - D_aB^a_k \approx 0 \ ,
 \qquad
   \phi^a_k \equiv 2 \ ( \varepsilon^{abc}R_{bc}^k
         - {\Lambda \over 12} B^a_k ) \approx 0 \ .
                                                            \eqno(10)
$$
 The Poisson brackets among them are given by
$$
\eqalign{
  & \{ \varphi_i({\bf x}), \varphi_j({\bf y})\} =
    - 2 \ \varepsilon _{ijk} \ \varphi_k({\bf x})
      \ \delta ^3 ({\bf x} - {\bf y}) \ ,
\qquad
  \{ \phi ^a _i({\bf x}), \phi^b_j({\bf y}) \} = 0  \ ,
\cr
  & \{ \varphi_i({\bf x}),\phi^a_j({\bf y}) \} =
     - 2 \ \varepsilon_{ijk} \ \phi^a_k({\bf x})
           \ \delta^3 ({\bf x} - {\bf y}) \ .}
                                                            \eqno(11)
$$
 All the constraints are of first class and the algebra is closed.
 The constraints $\varphi_k$ and $\phi^a_k$ generate the local-Lorentz
and $\theta_1^k$- transformations in (5) respectively.
 In this canonical formulation, the on-shell reducibility (6) appears as
a linear dependence of the constraints,
$$
    D_a \phi^a_k - {\Lambda \over 6} \varphi_k = 0 \ .
                                                           \eqno(12)
$$
 In the Dirac approach for quantization, one has to impose quantum
conditions to choose physical wave functional $\Psi$.
 In the topological case with $\Lambda \not= 0,$  these conditions can
be expressed using the constraints (10) in $\omega_a^k$ representation,
$$
\eqalign{
          & \varphi_k (\omega, \delta / \delta \omega) \Psi(\omega) =
            i D_a ( \delta / \delta \omega_a^k) \Psi(\omega) = 0  \ ,
\cr
          & \phi^a_k (\omega, \delta / \delta \omega) \Psi(\omega) =
            2 ( \varepsilon^{abc} R^k_{bc} + i {\Lambda \over 12} \,
            \delta / \delta \omega_a^k ) \Psi(\omega) = 0 \ .}
                                                             \eqno(13)
$$
 Since these equations are linear differential equations, we can
easily solve them to obtain the unique functional of $\omega_a^k$,
$$
    \Psi(\omega) = \exp( {6i \over \Lambda} I_{C-S} ) \ ,
\qquad
    I_{C-S} \equiv \int d^3x \varepsilon^{abc} \omega_a \cdot
       ( \partial_b \omega_c + {2 \over 3} ( \omega_b \times
          \omega_c ) ) \ ,
                                                             \eqno(14)
$$
where $I_{C-S}$ is the Chern-Simons term on the three-dimensional
boundary.
 This type of solution is also found in a different version of
topological gravity [7].
 The functional $\Psi(\omega)$ is the wave function of the universe
in four dimensional (Euclidean) topological gravity.
 It can also be considered as the BRST invariant vacuum, because it is
the unique representative annihilated by the BRST operator [1].  \par

\subsection{Einstein Gravity}

 On the other hand in the Einstein gravity ($\alpha \not=$ 0),
we have to solve the constraint equations,  which can be considered as
five linear equations for nine Lagrange multipliers $\Sigma _{a0}^k$ [2].
 The solution is expressed with four arbitrary variables $N^a$
(shift vector) and $\utilde{N}$ (lapse density of weight $-1$),
$$
  \Sigma_{a0}^k = -{1 \over 4} \   \varepsilon_{abc} [ N^b B^c_k
       + \utilde{N}( B^b \times B^c )^k ] \ .
                                                          \eqno(15)
$$
 Substituting this result for the canonical action (9),
we now have four constraints, together with $\varphi_k$ in (10),
which are associated with the Lagrange multipliers $N^a$ and
$\utilde{N}$,
$$
\eqalign{
          & C_a \equiv {1 \over 4} \  \varepsilon_{abc} B^b \cdot
                      \phi^c = B^b \cdot R_{ab} \approx 0 \ ,
\cr
          & C \equiv {1 \over 4} \  \varepsilon_{abc} ( B^a \times B^b)
                 \cdot \phi ^c = ( B^a \times B^b ) \cdot
                 ( R_{ab} - {\Lambda \over 24} \varepsilon_{abc}
                   B^c ) \approx 0 \ .}
                                                          \eqno(16)
$$
 We can identify $\varphi_k, C_a$ and $C$ with the independent first
class constraints corresponding to the generators of the local-Lorentz
transformation, spatial diffeomorphism and temporal diffeomorphism
respectively.
\par
 An important observation is that all the constraints in the Einstein
Gravity are linear combinations of those in the topological version.
 Especially four diffeomorphism generators $C_a$ and $C$ in (16) are
linearly dependent on nine `new-type' generators $\phi^a_k$ in (10).
 We see that  $\Psi(\omega)$ in (14) becomes a special solution of all
quantum constraints in the Einstein gravity if the operator ordering is
arranged as in (16).
 This ordering is consistent with the commutation relations among the
constraint operators [8].
 This $\Psi(\omega)$ is nothing but the Euclidean version of the wave
functional discovered by Kodama [9,10].
 Therefore the physical space of the topological gravity is contained
in that of the Einstein gravity.

\subsection{New System}

 With the vector field $\eta^k$, the action becomes in canonical form,
$$
 S =
   \int dt \int d^3x [ \dot \omega_a \cdot B^a - \omega _0 \cdot \varphi
        - \Sigma_{a0} \cdot \phi^a
        -2\alpha \psi_{kl} {\hat \Sigma}^k_{a0} {\hat B}^a_l] \ ,
                                                               \eqno(17)
$$
where ${\hat B}^a_k \equiv \epsilon^{abc} {\hat \Sigma}_{bc}^k$
is the spatial components of the 2-form ${\hat \Sigma^k}$.
 Again we have to solve the constraint equation derived by varying (17)
with respect to $\psi_{kl}$.
 The solution is expressed by using four arbitrary variables $N^a$
and $\utilde{N}$,
$$
  \Sigma_{a0}^k = -{1 \over 4} \   \epsilon_{abc} [ N^b {\hat B}^c_k
       + \utilde{N}( {\hat B}^b \times {\hat B}^c )^k ]
       - {1 \over 2}{\dot \eta}_a^k -(\omega_0 \times \eta_a)^k
       + {1 \over 2}{\hat D}_a \eta^k_0  \ .
                                                          \eqno(18)
$$
 Substituting this result for the canonical action (17), we get four
sets of constraints:
$$
\eqalign{
 & {\hat \varphi} _k \equiv - D_aB^a_k -2 (\eta_a \times
                            {}^{\eta}\pi^a)^k
                     \approx 0 \ ,
\cr
 & {\hat \phi}^a_k \equiv 2 \ ( \epsilon^{abc}R_{bc}^k
                   - {\Lambda \over 12} B^a_k ) -2 \, {{}^{\eta}\pi}^a_k
                                       \approx 0 \ ,
\cr
 & H_a \equiv {1 \over 4} \  \epsilon_{abc} {\hat B}^b \cdot
                   ({\hat \phi^c} + 2 \, {{}^{\eta} \pi^c}) \approx 0,
\cr
 & {\cal H} \equiv {1 \over 4} \  \epsilon_{abc} ({\hat B}^a \times
            {\hat B}^b) \cdot ({\hat \phi^c} + 2 \, {{}^{\eta} \pi^c})
             \approx 0 \ .
}
                                                           \eqno(19)
$$
 The fields ${{}^{\eta}\pi}^a_k$ are the conjugate momenta of the
spatial components of $\eta_a^k$.
 Next we redefine the constraint $H_a$ as
$$
 {\cal H}_a \equiv 2 H_a + \omega_a \cdot {\hat \varphi}
              - {1 \over 2} \epsilon_{abc}
             ({\hat B}^b -B^b) \cdot {\hat \phi^c} \ .
                                                          \eqno(20)
$$
 The new constraint ${\cal H}_a$ generates the spatial diffeomorphism.
 The non-zero Poisson brackets among the constraints are given by
$$
\eqalign{ &
   \{ {\hat \varphi}[g_1], {\hat \varphi}[g_2] \} =
    - 2 \, {\hat \varphi}[(g_1 \times g_2)] \ ,
\qquad
    \{ {\hat \varphi}[g], {\hat \phi}^a [h_a] \} =
     - 2 \, {\hat \phi}^a [(g \times h_a)] \ ,
\cr
      &   \{{\cal H}[{\utilde N}], {\cal H}[{\utilde M}] \}
   =  {\cal H}_a [L^a] - {\hat \varphi} [L^a \omega_a ]
     + {1 \over 2} {\hat \phi}^a
       [\epsilon_{abc} L^b ({\hat B}^c - B^c )] \ ,
}
                                                            \eqno(21)
$$
where ${\hat \varphi}[g_1] \equiv \int d^3x g_1^k {\hat \varphi}_k, \
{\hat \phi}^a [h_a] \equiv \int d^3x h_a^k {\hat \phi}^a_k $  $\, $
 $L^a \equiv {\hat B}^a \cdot {\hat B}^b ({\utilde M} \partial_b
{\utilde N} - {\utilde N} \partial_b {\utilde M})$.
 All the constraints in the system are of first class.
 Among them, ${\hat \varphi}_k$ and ${\hat \phi}^a_k$ are identified
with the generators of the local-Lorentz and $\theta_1^k$-
transformations in (8) respectively.
 The constraint ${\cal H}$ generates temporal diffeomorphism while
${\cal H}_a$ the spatial one.\par

 As in the previous cases, quantum conditions are
$$
\eqalign{
          & {\hat \varphi}_k (\omega, \delta / \delta \omega ,
             \eta, \delta / \delta \eta ) \Psi(\omega , \eta) =
              0  \ ,
\qquad
           {\hat \phi}^a_k (\omega, \delta / \delta \omega ,
             \eta, \delta / \delta \eta ) \Psi(\omega , \eta) =
              0  \ ,
\cr
          & {\cal H}_a (\omega, \delta / \delta \omega ,
             \eta, \delta / \delta \eta ) \Psi(\omega , \eta) =
              0  \ ,
\qquad
           {\cal H} (\omega, \delta / \delta \omega ,
             \eta, \delta / \delta \eta ) \Psi(\omega , \eta) =
              0  \ .
}
                                                             \eqno(22)
$$
 These linear differential equations are solved as the following [11]:
$$
\eqalign{
    \Psi(\omega, \eta)
       & = \exp[ {6i \over \Lambda} I_{C-S}
               + \beta (I_{C-S} - {\Lambda \over 6} I_{New})] \, ,
\cr
   I_{New} \equiv
   &  \int d^3x {\varepsilon}^{abc} {\eta}_a \cdot R_{bc} \,
       - {\Lambda \over 24} \int d^3x {\varepsilon}^{abc} {\eta}_a
                       \cdot D_b {\eta}_c
\cr
   &         + {4 \over 3} \, ({\Lambda \over 24})^2 \int d^3x
                   {\varepsilon}^{abc} {\eta}_a \cdot ( {\eta}_b
                      \times {\eta}_c ) \, ,
}
                                                              \eqno(23)
$$
where $\beta$ is a constant parameter. If we choose it as
$ - {6 \over \Lambda} i $, the wave function of the Universe becomes
$$
        \Psi(\omega, \eta) = \exp( i I_{New} ) \, .
                                                              \eqno(24)
$$
This solution is suitable for both $ \Lambda = 0 $ and
$ \Lambda \not= 0 $ cases.

\chapter{Summary}   %

 We have clarified the relation between the 2-form Einstein gravity
 and its topological version.
 The physical space of the topological version is contained in that of
the Einstein gravity. \par
 Moreover the vector field $\eta^k$ is introduced into 2-form Einstein
gravity to restore the large symmetry of its topological version.
 Since this new model has the modes of the gravitational wave,
it is equivalent to the Einstein gravity.
 It may be a good strategy in quantum gravity to stydy models
with large symmetry, in addition to local Lorentz and diffeomorphisms.
\par
 We have obtained the wave function of the universe for each model.
\vskip 2cm

\subsection{Acknowledgements}

 I am grateful to Prof. H.Y. Lee and Dr. T. Ueno for the collaboration.

\baselineskip=24pt

\vfill

\eject



\REF\LNU{H.Y. Lee, A. Nakamichi, and T. Ueno,
 {\it `Topological 2-form Gravity in Four Dimensions'},
Phys.$\>$Rev.$\>$D47 (1993) 1563. }

\REF\SEECDJM{R. Capovilla, J. Dell, T. Jacobson and L. Mason,  \nextline
Class.$\>$Quantum$\>$Grav.$\>$8 (1991) 41.}

\REF\SEEA{A. Ashtekar, Phys.$\>$Rev.$\>$D36 (1987) 1587; {\it `New
          Perspectives in Canonical Gravity'}, (Bibliopolous, Naples,
          Italy, 1988).}

\REF\SEEAN{A. Nakamichi and T. Ueno,
{\it `New Vector Field and BRST Charges in 2-form Einstein Gravity'},
 TIT/HEP-206/COSMO-24, February 1993.}

\REF\SEEPR{R. Penrose and W. Rindler, {\it `Spinors and Space-time'}
Vols. I, II (Cambridge University Press, Cambridge, 1984).}

\REF\SEEJS{T. Jacobson and L. Smolin, Class.$\>$Quantum$\>$Grav.$\>$5
(1988) 583.}

\REF\SEEHO{G.T. Horowitz, Commun. Math. Phys. 125 (1989) 417.}

\REF\SEEAP{A. Ashtekar, Phys. Rev. Lett. 57 (1986) 2244.}

\REF\SEEK{H. Kodama, Phys.$\>$Rev.$\>$D42 (1990) 2548.}

\REF\SEEI{H. Ikemori, in {\it `Proceeding of the Workshop on Quantum
             Gravity and Topology'}, edited by I. Oda (Institute for
             Nuclear Study, University of Tokyo, 1991). }

\REF\SEETU{A. Nakamichi and T. Ueno, in preparation.}


\refout

\end